\documentclass[lettersize,journal]{IEEEtran}

\usepackage{amsmath,amsfonts}
\usepackage{algorithm}
\usepackage{algpseudocode}
\usepackage{array}
\usepackage{textcomp}
\usepackage{stfloats}
\usepackage{url}
\usepackage{verbatim}
\usepackage{graphicx}
\usepackage{cite}
\usepackage{subcaption}
\usepackage{hyperref}
\usepackage{tabularx}
\begin{document}

\title{A dataset and model for auditory scene recognition for hearing devices: AHEAD-DS and OpenYAMNet}

\author{Henry Zhong, Jörg M. Buchholz, Simon Carlile, Julian Maclaren, Richard Lyon%
\thanks{Henry Zhong (e-mail: henry.zhong@mq.edu.au), Jörg M. Buchholz. Australian Hearing Hub, Macquarie University, Sydney, Australia.}
\thanks{Simon Carlile (e-mail: scarlile@google.com), Julian Maclaren, Richard Lyon. Google Research Australia, Sydney, Australia.}
}

\maketitle

\begin{abstract}
Scene recognition is important for hearing devices, however; this is challenging, in part because of the limitations of existing datasets. Datasets often lack public accessibility, completeness, or audiologically relevant labels, hindering systematic comparison of machine learning models. Deploying such models on resource-constrained edge devices presents another challenge. The proposed solution is two-fold, a repack and refinement of several open source datasets to create AHEAD-DS, a dataset designed for auditory scene recognition for hearing devices, and introduce OpenYAMNet, a sound recognition model. AHEAD-DS aims to provide a standardised, publicly available dataset with consistent labels relevant to hearing aids, facilitating model comparison. OpenYAMNet is designed for deployment on edge devices like smartphones connected to hearing devices, such as hearing aids and wireless earphones with hearing aid functionality, serving as a baseline model for sound-based scene recognition. OpenYAMNet achieved a mean average precision of 0.86 and accuracy of 0.93 on the testing set of AHEAD-DS across fourteen categories relevant to auditory scene recognition. Real-time sound-based scene recognition capabilities were demonstrated on edge devices by deploying OpenYAMNet to an Android smartphone. Even with a 2018 Google Pixel 3, a phone with modest specifications, the model processes audio with approximately 50ms of latency to load the model, and an approximate linear increase of 30ms per 1 second of audio. The project website with links to code, data, and models. \href{https://github.com/Australian-Future-Hearing-Initiative}{https://github.com/Australian-Future-Hearing-Initiative}
\end{abstract}


\section{Introduction}
State of the art hearing devices incorporate scene recognition in order to identify the auditory environment \cite{vivek2020acoustic}. Scene recognition can be used to trigger distinct hearing device processing strategies \cite{cauchi2018hardware}, such as wind reduction, beamforming, and music mode. Processing strategies improve a hearing device user's experience by enhancing salient sounds while suppressing undesirable background noise.

Several machine learning (ML) models have been proposed for scene recognition \cite{nigro2024trends}. It is important that models can run locally to preserve privacy, by avoiding the transmission of sensitive data and to minimise the cost of data transmission. Hearing devices like hearing aids, wireless earphones with hearing aid functionality, and smartphones that operate in tandem with hearing devices are edge devices \cite{garcia2023analysing}. Porting ML models to edge devices is challenging because such devices are resource constrained \cite{chen2020deep} and have limited data format compatibility \cite{murshed2021machine}.

The challenge of scene recognition is further compounded by the gap in publicly available standardised audio scene recognition datasets. Existing datasets are either in-accessible, incomplete, or not annotated with labels relevant to hearing device listening situations. These limitations prevent the systematic evaluation of ML models for scene recognition.

The motivation for this paper is to address the aforementioned challenges to improve scene recognition. In doing so, allow better application of processing strategies to improve the hearing experience for users of hearing devices. This paper makes the following three contributions. Firstly, a description of a sound dataset named \emph{Another HEaring AiD scenes Data Set} (AHEAD-DS). This dataset was created for the purpose of scene recognition of the auditory environment, which allows comparison of models using a standardised dataset with consistent labels. The dataset is repacked and refined from several other open source datasets, released under a permissive licence, in a ready-to-use state, divided into training/validation/testing sets and annotated with the ground truth. Secondly, a description of a ML model called OpenYAMNet for sound based scene recognition. The model can act as a baseline model for further development of sound based scene recognition models. The model is released under a permissive licence, including model definitions, training, testing code, and code to convert it into a format deployable to edge devices. Finally, benchmark results of an implementation of the OpenYAMNet ML model, trained using AHEAD-DS, including precision, recall and inference speed. The ML model was deployed to a smartphone to demonstrate the model's functionality on an edge device.

The remainder of this paper is organised as follows. The remainder of this section presents related work through summaries of existing datasets and ML models. The section that follows presents the methods used to create AHEAD-DS and the OpenYAMNet model. The section after presents test results. The penultimate section discusses the findings of the test results, compares against existing work, and describes the possible future direction of the research. The final section draws some conclusions.

\subsection{Existing Audio Scene Recognition Datasets for Hearing Devices}
Several existing datasets were collected as part of studies relevant to auditory scene recognition, a summary is shown in Table \ref{tab:datasummary}. The table summarises advantages and issues for individual datasets.

The main challenges of existing audio scene recognition datasets for hearing devices fall into three categories. Public accessibility, completeness, and labelling relevance. Accessibility was a problem for datasets which were immediately discarded after processing, or under a proprietary and restrictive licence. The issue of completeness stems from datasets being a mix of secondary sources, but the mixing procedure and specific subset from secondary sources are unknown.

Several large general-purpose datasets exist for sound recognition. Datasets such as AudioSet \cite{gemmeke2017audio}, UrbanSound8K \cite{salamon2014dataset}, ESC50 \cite{piczak2015esc}, and TAU scenes \cite{wang2021curated} provide extensive training material for ML models. However, the existing labels are either too specific or too general for the objective of auditory scene recognition for hearing devices. For example, AudioSet recordings can be labelled bus or truck. This level of granular sound recognition is unnecessary for a hearing device, it is only necessary to recognise and respond to broad categories like motor vehicle noises to improve the user's experience. However, this data can still be used through transfer learning, or the labels can be remapped to a more general set.

Conversely, TAU scenes provide location-based recordings which are too general, such as at airports and metro stations. The sound of a motor vehicle will be similarly disruptive irrespective of whether the sound occurs at either location. However, labels could be modified and made more specific, but the modification would likely require listening to and re-annotating the recordings.

Another challenge is the use of different labels across datasets. It is difficult to directly compare ML models trained on each dataset and makes it difficult to combine the datasets into a single, more comprehensive one.

\begin{table*}
\centering
\captionsetup{justification=centering}
\caption{This table summarises relevant datasets for auditory scene recognition for hearing devices.}
\begin{tabularx}{\textwidth}{p{0.1\textwidth}p{0.25\textwidth}p{0.1\textwidth}p{0.45\textwidth}}
\hline
\textbf{Study} & \textbf{Labels} & \textbf{Data Licence} & \textbf{Notes} \\
\hline
\cite{humes2018acoustic} & Quiet, Speech moderate, Speech high, Noise moderate, Noise high, Speech in noise moderate, Speech in noise high & Audio data not saved & Sound data processed on-the-fly. The classification results were recorded but sound was not. The classification system was a proprietary part of the hearing aids used in the study. \\
\cite{huwel2020hearing} & Cocktail party, Interfering speakers, In traffic, Speech in traffic, In vehicle, Speech in vehicle, Music, Speech in music, Quiet indoors, Speech in quiet indoors, Reverberant env., Speech in reverb. env. , Wind turbulence, Speech in wind turb. & Open source & Sound dataset derived from CHiME 5 \cite{barker18_interspeech}, GTZAN \cite{tzanetakis2002musical}, the environment sounds recorded by authors of the study are released publicly. Though the number of samples released is not consistent with the description in the study. The code for mixing speech and environment sounds is missing. Dataset is named Hearing Aid Research Data Set for Acoustic Environment Recognition (HEAR-DS). \\
\cite{vivek2020acoustic} & Music, Noise, Speech with noise, Speech, Silence & Mix of open source and proprietary & Sound dataset derived from freesound.org, Demand database \cite{thiemann2013diverse}, CSTR VCTK corpus \cite{yamagishi2019cstr}, Million Song Dataset \cite{bertin2011million}, LibriSpeech ASR corpus \cite{panayotov2015librispeech} and YouTube. YouTube source is proprietary. The data split and any preprocessing parameters are unknown. \\
\cite{fan2020deep} & Bus, Subway, Street, Indoor, Car & Proprietary & Sound dataset recorded by study authors, not released publicly. \\
\cite{kuebert2021daily} & Transportation, Physical, Basics, Social, Listening & Audio data not saved & Sound features and acceleration features are computed and recorded, but sound was not recorded. \\
\cite{christensen2021everyday} & Quiet, Speech, Speech in noise, Noise & Audio data not saved & Sound data was collected and processed on-the-fly and not saved. \\
\cite{fabry2021improving} & Speech in restaurant, Speech in car, Speech in large room & Proprietary & The test results and labels come from a separate study \cite{walsh20202} provided by a hearing aid manufacturer. The data was not publicly available. \\
\cite{hayes2021environmental} & Music, Single talker food court, One talker on subway platform, Conversation in quiet & Proprietary & Labels from five hearing aid manufacturers are mapped to the labels used in the study. Sound was recorded by the authors and not released publicly. \\
\cite{ting2021environmental} & Vehicle engine, Machine tools, Household appliance, Nature, Human speech & Open source & Sound dataset derived from UrbanSound8K \cite{salamon2014dataset} and ESC50 \cite{piczak2015esc}. Remapped to relevant labels. An unknown amount of data from the two source datasets was removed. Data augmentation is used but it was not described in sufficient detail to be reproducible. \\
\cite{yellamsetty2021comparison} & Speech in quiet, Speech in noise, Noise, Music & Mix of open source and proprietary & Sound dataset derived from proprietary Phonak Sound CD 2, self recordings on the London tube subway, and a song from royalty free freesound.org. \\
\cite{wang2021curated} & Airport, Bus, Metro, Metro station, Park, Public square, Shopping mall, Street pedestrian, Street traffic, Tram & Open source & The TAU urban audio visual scenes datasets were recorded in several European cities." \\
\hline
\end{tabularx}
\label{tab:datasummary}
\end{table*}

\subsection{Existing Sound Recognition Models}
Several existing sound recognition models exist, the three presented in this section are Yet Another Mobile Network (YAMNet) \cite{plakal2019yamnet}, Perch \cite{ghani2023global}, and Audio Spectrogram Transformer (AST) \cite{gong21b_interspeech}. The three representative models were selected as the models provide a good cross-section of common deep learning designs, many other ML models are variations of the three architectures \cite{zaman2023survey}.

AST and YAMNet were pre-trained, validated, and tested using the AudioSet dataset, while Perch was pre-trained using XenoCanto \cite{vellinga2015xeno}. The extensive datasets used for pre-training makes any of the models suitable for fine tuning via transfer learning. AST and YAMNet use log mel spectrograms as features, while Perch uses a per-channel energy normalised mel spectrogram. The spectrograms are represented as two-dimensional inputs and processed using neural networks designed for image recognition.

YAMNet's implementation is light weight at 3.7 million parameters compared to Perch and AST, which use 7.8 and 86 million parameters respectively. This makes YAMNet a good match for edge computing. However, YAMNet's implementation is open weight but not fully open source, since only model definitions and weights have been released. The training code for YAMNet is missing from the code repositories, while Perch and AST are fully open source.

\section{Methods}
This section outlines the methodology used to create the AHEAD-DS dataset and the OpenYAMNet model. The procedures for data curation, preprocessing, and labelling for AHEAD-DS are detailed, followed by a description of the OpenYAMNet model architecture, training, and deployment on edge devices.

\subsection{Another HEaring AiD scenes Data Set (AHEAD-DS)}
AHEAD-DS repackages and refines the open source datasets HEAR-DS and CHiME 6 Dev \cite{watanabe2020chime}. A summary of the dataset is shown in Table \ref{tab:aheadds}. There are a total of 9968 recordings, 10s per recording, across 14 different labels. Each instance is single-channel, 16-bit integers and sampled at 16 kHz.

AHEAD-DS follows the design of HEAR-DS and employs an identical set of 14 labels. This set comprises six environmental sound categories and an additional six categories representing speech instances within these respective environments.

The sounds from each environment for AHEAD-DS were taken from HEAR-DS. The authors of HEAR-DS recorded each scene using a dummy head. The head was equipped with a synthetic ear, the GRAS KB1065/1066 Pinnae. An Audifon microphone was placed in each ear of the dummy head in an \emph{in-the-canal} position. A Focusrite Scarlett 18i6 sound interface was used for analogue to digital conversion. All recordings were initially captured as 32-bit floating-point values sampled at 48 kHz, and later downsampled to a 16 kHz sampling rate with 16-bit integer quantisation. There were a total of 5929 clips of 10s, from 6 different environments.

Music samples from 10 genres were taken from HEAR-DS. The authors of HEAR-DS obtained the music from GTZAN \cite{tzanetakis2002musical}. All music samples were 10s long, 16-bit integers and sampled at 16 kHz. There were a total of 2992 clips of 10s of music.

CHiME 6 Dev was used for speech samples in AHEAD-DS instead of CHiME 5, which was used in HEAR-DS. Both contain the same speech recordings, but the start and end times of all recordings have been corrected in CHiME 6 Dev to ensure all recordings are synchronised. CHiME 6 dev is a subset of CHiME 6, it is smaller than the whole dataset but sufficiently long for mixing with environment sounds to provide all the \emph{speech in $\langle environment \rangle$} clips and to produce \emph{interfering speakers}. The \emph{interfering speakers} clips were derived entirely from CHiME 6 dev.

The authors of CHiME 6 dev recorded the speech in two sessions. Each session in two different home environments consisting of an open space with dining, kitchen, and lounge areas. A total 44 sound files from CHiME 6 Dev were used in AHEAD-DS, 24 recordings from the first environment and 20 in the second. Each session consisted of four people and would begin by having participants take turns to introduce the participants via speaking and then read out a short passage. Once the introduction is complete, participants engage in casual conversation as the participants prepare and share a meal together.

The authors of CHiME 6 dev made the recordings by placing six Microsoft Kinects throughout the home in the first session and five in the second. However, the Kinect version is unspecified. Each Kinect recorded four sound streams simultaneously, using 16-bit integers, sampled at 16 kHz.

In CHiME 6 additional speech was recorded using body-worn microphones, but only the recordings from the Kinect were used. The recordings from body-worn microphones lacked acoustic realism as the recordings sounded like clean studio recordings.

The label \emph{interfering speakers} contains only speech taken from CHiME 6 and is not mixed with environment sounds. The label represents a scenario with multiple speakers where the speech is intelligible. The label \emph{cocktail party} contains multiple speakers but the speech is unintelligible.

\begin{table*}
\centering
\captionsetup{justification=centering}
\caption{This table summarises the contents of AHEAD-DS. Each recording is 10s, single channel, 16-bit integers and sampled at 16 kHz.}
\label{tab:aheadds}
\begin{tabularx}{0.9\textwidth}{p{0.3\textwidth}p{0.12\textwidth}p{0.12\textwidth}p{0.12\textwidth}p{0.12\textwidth}}
\hline
\textbf{Label} & \textbf{Training} & \textbf{Validation} & \textbf{Testing} & \textbf{All} \\
\hline
cocktail\_party & 934 & 134 & 266 & 1334 \\
interfering\_speakers & 733 & 105 & 209 & 1047 \\
in\_traffic & 370 & 53 & 105 & 528 \\
in\_vehicle & 409 & 59 & 116 & 584 \\
music & 1047 & 150 & 299 & 1496 \\
quiet\_indoors & 368 & 53 & 104 & 525 \\
reverberant\_environment & 156 & 22 & 44 & 222 \\
wind\_turbulence & 307 & 44 & 88 & 439 \\
speech\_in\_traffic & 370 & 53 & 105 & 528 \\
speech\_in\_vehicle & 409 & 59 & 116 & 584 \\
speech\_in\_music & 1047 & 150 & 299 & 1496 \\
speech\_in\_quiet\_indoors & 368 & 53 & 104 & 525 \\
speech\_in\_reverb\_env & 155 & 22 & 44 & 221 \\
speech\_in\_wind\_turbulence & 307 & 44 & 88 & 439 \\
\textbf{Total} & \textbf{6980} & \textbf{1001} & \textbf{1987} & \textbf{9968} \\
\hline
\end{tabularx}
\end{table*}

\subsection{AHEAD-DS Data Processing Workflow}
\begin{figure*}
\centering
\captionsetup{justification=centering}
\includegraphics[width=1.0\textwidth]{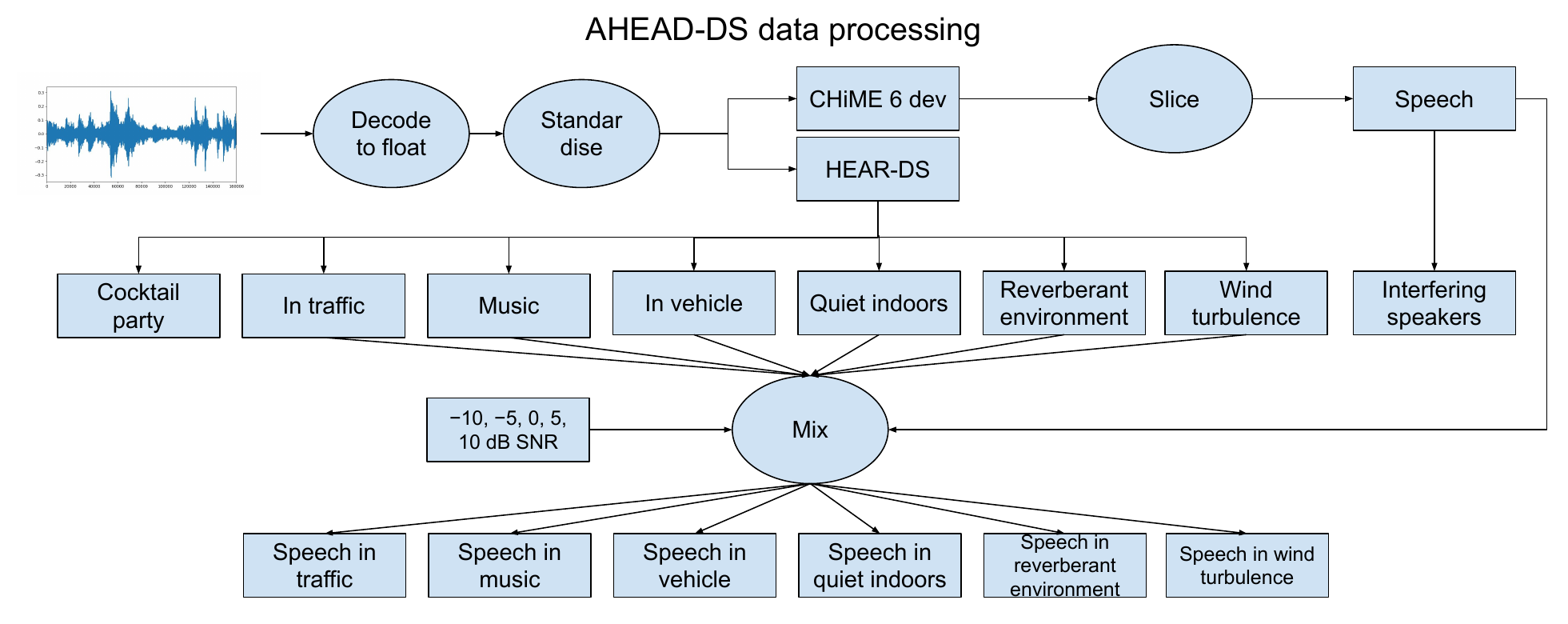}
\caption{A data flow diagram showing the data processing workflow.}
\label{fig:dataprocessing}
\end{figure*}

The workflow to create AHEAD-DS is shown in Figure \ref{fig:dataprocessing}. The sounds sourced from HEAR-DS and CHiME 6 Dev had been encoded in Waveform Audio File Format (WAV) as single-channel, 16-bit integers, and sampled at 16 kHz using several different equipment configurations. AHEAD-DS retained this format. For all level adjustments and mixing, the audio files were converted into 32-bit floating-point format during processing. The values were then clipped to the values of $int16_{max}$ and $int16_{min}$ to avoid quantisation errors, then converted back to 16-bit integers, before being saved to disk.

The levels of CHiME 6 dev and \emph{music} were high as the levels were optimised for playback. The signal levels of the other recordings were much lower, and not optimised for playback. The levels were standardised before mixing the clips. For everything except CHiME 6 Dev and \emph{music}, the audio was standardised by dividing by the RMS of \emph{cocktail party} and multiplying by the RMS of CHiME 6 Dev. For \emph{music} the audio was standardised by dividing by the RMS of \emph{music} and multiplying by the RMS of CHiME 6 Dev.

These adjustments ensure that the levels of \emph{cocktail party} and \emph{music} were approximately similar to conversations in CHiME 6 Dev. Recordings using new hardware configurations must be standardised, this only needs to be done once per device, to ensure recordings of conversations using the new device have similar levels to CHiME 6 Dev.

The first second from each recording in CHiME 6 dev is discarded, as the first second contains a loud beep to signal the start of the recording. The next 1100s were extracted from each file in 10s intervals. This generated 110 clips per file for a total of 4840 clips of speech.

For every environment from HEAR-DS, half the clips were mixed with speech, the other half are kept unmixed to represent the environment label. There are six environment labels which need to be mixed with speech. Of the 4840 speech clips taken from CHiME 6 Dev, 3793 clips were mixed with environmental sounds. 1047 clips were assigned to the label \emph{interfering speakers} without further modifications.

Every five speech and environment sound file pairs are mixed at $-10, -5, 0, 5, 10$ dB signal-to-noise ratio (SNR). The mixing function is shown in Algorithm \ref{alg:mix}. The reason to mix speech at $-10$ dB SNR was to allow for the possibility to detect speech, though not intelligibility, even in edge cases that would challenge a human with good hearing. When speech and environment sounds were mixed, the levels of the quieter sound were boosted until it had equal RMS to the louder sound, this procedure occurs on top of standardisation. During mixing, speech was used as the signal and the environment sounds were used as noise. The speech levels were adjusted to the required SNR. The sounds of \emph{cocktail party} and \emph{interfering speakers} were used without mixing. No speech clip was used more than once.

The reason to boost the quieter of \emph{speech} or \emph{environment sound} is to avoid the following situation. When \emph{speech} was mixed at $-10$ dB SNR with quiet \emph{environment sounds} such as \emph{quiet indoors}, speech was imperceptible. This caused sound clips with imperceptible speech being labelled with a \emph{speech in environment} label e.g. \emph{speech in quiet indoors}, which is detrimental to model performance. The level boost ensures sound clips match the label.

After standardisation and mixing, sounds are divided into training, validation, and testing sets using a percentage split of 70, 10, and 20 respectively.

\begin{algorithm}
\caption{Pseudocode to mix speech with environment sound.}
\label{alg:mix}
\begin{algorithmic}[1]
\Function{Mix}{$speech, env, snr$}
    \State $speech_{rms} \gets \text{RMS}(speech)$
    \State $env_{rms} \gets \text{RMS}(env)$
    \State \Comment{Prevent division by zero}
    \If{$speech_{rms} == 0$}
        \State $speech_{rms} \gets 1e-6$
    \EndIf
    \If{$env_{rms} == 0$}
        \State $env_{rms} \gets 1e-6$
    \EndIf
    \State \Comment{Normalise amplitudes}
    \If{$speech_{rms} > env_{rms}$}
        \State $speech_{boosted} \gets speech$
        \State $env_{boosted} \gets env \times \frac{speech_{rms}}{env_{rms}}$
    \Else
        \State $speech_{boosted} \gets speech \times \frac{env_{rms}}{speech_{rms}}$
        \State $env_{boosted} \gets env$
    \EndIf
    \State \Comment{Calculate gain from SNR}
    \State $gain_{factor} \gets 10^{\left(\frac{snr}{20}\right)}$
    \State \Comment{Apply gain and mix}
    \State $speech_{gained} \gets speech_{boosted} \times gain_{factor}$
    \State $mixed \gets speech_{gained} + env_{boosted}$
    \State \textbf{return} $mixed$
\EndFunction
\end{algorithmic}
\end{algorithm}

\subsection{OpenYAMNet Model}
\begin{figure}
\centering
\captionsetup{justification=centering}
\includegraphics[width=0.45\textwidth]{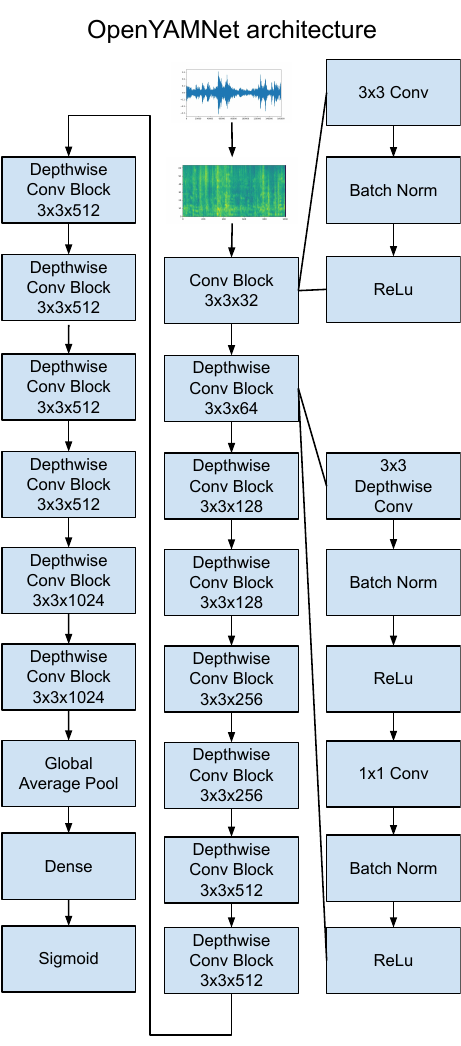}
\caption{The architecture of OpenYAMNet.}
\label{fig:yamnetp}
\end{figure}

The architecture diagram of OpenYAMNet is shown in Figure \ref{fig:yamnetp}. OpenYAMNet is based on YAMNet \cite{plakal2019yamnet} with the same internal architecture. A complete workflow for OpenYAMNet was created, as the original YAMNet was released with model definitions and weights but missing other components such as training code. All code has been released, making the implementation fully open source. For OpenYAMNet the code consists of training code, code to load weights, code for training via transfer learning, testing code, and conversion code to downsampling the model for deployment to edge computing platforms.

OpenYAMNet takes the waveform as input. A log mel spectrogram is computed from the waveform. The spectrogram is treated as two dimension input and processed using MobileNet \cite{howard2017mobilenets}. MobileNet starts with a convolutional block and a series of 13 blocks containing a depthwise separable convolution, batch normalisation, ReLU activation, convolution, batch normalisation and ReLU activation. There is a penultimate global average pooling layer. The final layer is a fully connected layer containing sigmoid activation of length equal to the number of class labels.

OpenYAMNet slices the waveform into windows of length 960ms with a 480 millisecond overlap, therefore, a 10 second clip will be split into 20 windows. Zero padding is added to ensure the input waveform is of sufficient length. Scene recognition is performed on every 960 millisecond window and each window is treated separately. The 960 millisecond window length is a hyperparameter which is fixed by the use of transfer learning.

\subsection{OpenYAMNet Training Workflow}
OpenYAMNet weights were initialised using transfer learning \cite{zhuang2020comprehensive}, taken from YAMNet trained on AudioSet. Training occurred for a maximum of 100 epochs. Focal loss \cite{lin2017focal} with $\alpha=0.25$ and $\gamma=2$ was used as the loss function. An Adam optimiser \cite{kingma2015adam} was used, with an initial learning rate of 0.00001. A decay-on-plateau learning rate scheduler was used. After 3 epochs, if performance on the validation set did not improve, the learning rate decays by multiplying by 0.5. After 6 epochs, if performance on the validation set did not improve, the training cycle was terminated. The order of the training data was shuffled at the end of every epoch.

Label smoothing \cite{muller2019does} with a smoothing factor of 0.1 was applied as a regularisation function. Data augmentation \cite{shorten2019survey} was also applied as a regularisation technique. Three augmentation procedures were applied, each independently with a probability of fifty percent: a gain of between -6 to 6 dB, noise to perturb the data, from a uniform distribution within range -0.003 to 0.003, and the length of the clip is shrunk or stretched between 0.9 and 1.1, using the Fourier resampler \cite{OppenheimSchaferBuck}.

The window length of 960 milliseconds and overlap of 480 milliseconds causes each 10 second audio clip in AHEAD-DS to be divided into 20 windows. The dataset contains 6,980 training, 1,001 validation, and 1,987 testing clips. A total of 139,600 windows are used for training, 20,020 for validation, and 39,740 for testing.

\subsection{Software Environment}
OpenYAMNet was developed using Python and TensorFlow. For technical details, links to the code are provided in the \emph{Data availability statement}. The models were trained using only the central processing unit (CPU). It was not possible to get graphics processing unit (GPU) acceleration due to compatibility issues with TensorFlow, however, GPU use would reduce training time significantly.

In order to optimise OpenYAMNet for deployment on edge devices, the full-precision model was converted to a half-precision 16-bit floating-point model. This was done using TensorFlow Lite. The model was deployed through a phone application \cite{prismpaper} written using Flutter. The edge computing test system was a Google Pixel 3 running Android.

\section{Results}
This section outlines the results of testing OpenYAMNet using AHEAD-DS. The analysis includes the model's performance on precision, recall, and accuracy. It also compares the impact of various hyperparameters, evaluates the processing time on an edge platform, and benchmarks the model against several other ML models.

\subsection{Precision, Recall, and Accuracy}
\begin{figure*}
\centering
\captionsetup{justification=centering}

\begin{subfigure}[t]{0.6\textwidth}
\includegraphics[width=\textwidth]{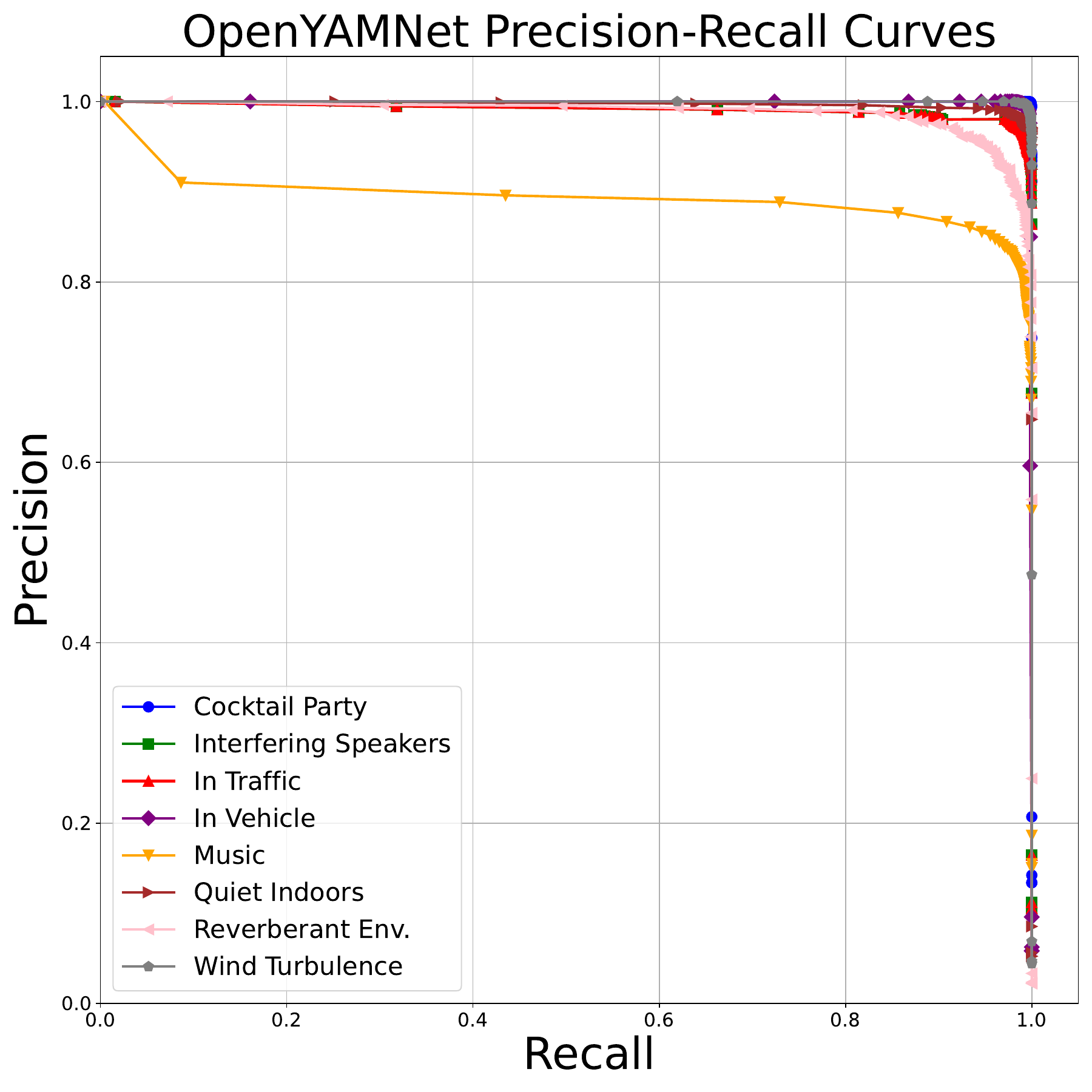}
\caption{OpenYAMNet precision and recall curves for cocktail party, interfering speaker, and several environments.}
\end{subfigure}

\begin{subfigure}[t]{0.6\textwidth}
\includegraphics[width=\textwidth]{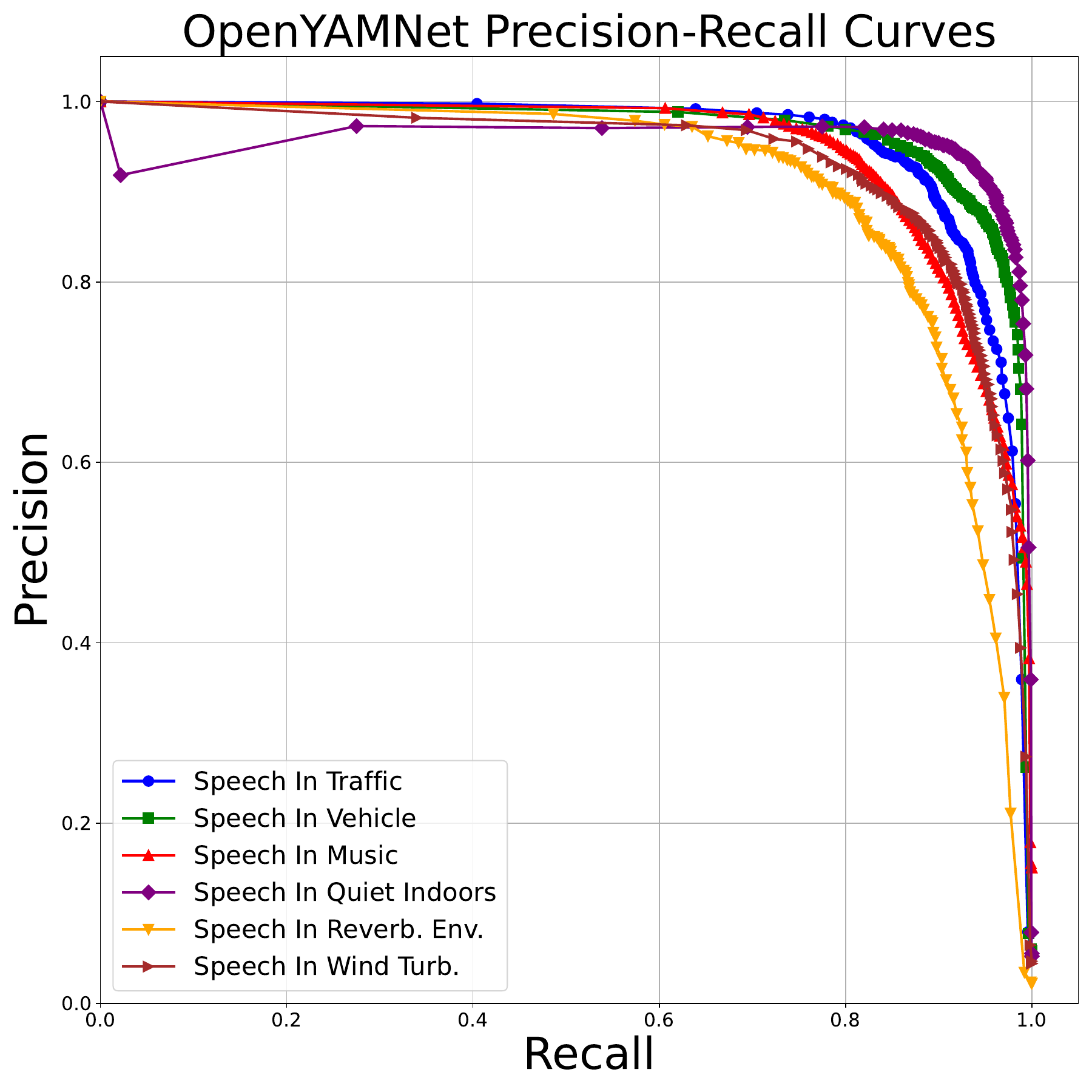}
\caption{OpenYAMNet precision and recall curves for speech in several environments.}
\end{subfigure}

\caption{OpenYAMNet precision and recall curves. A mAP of 0.86 was achieved on the test set. Each 960 millisecond window contributed a prediction outcome. A larger area under the curve indicates better performance.}
\label{fig:precisionrecall}
\end{figure*}

\begin{figure*}
\centering
\captionsetup{justification=centering}
\includegraphics[width=1.0\textwidth]{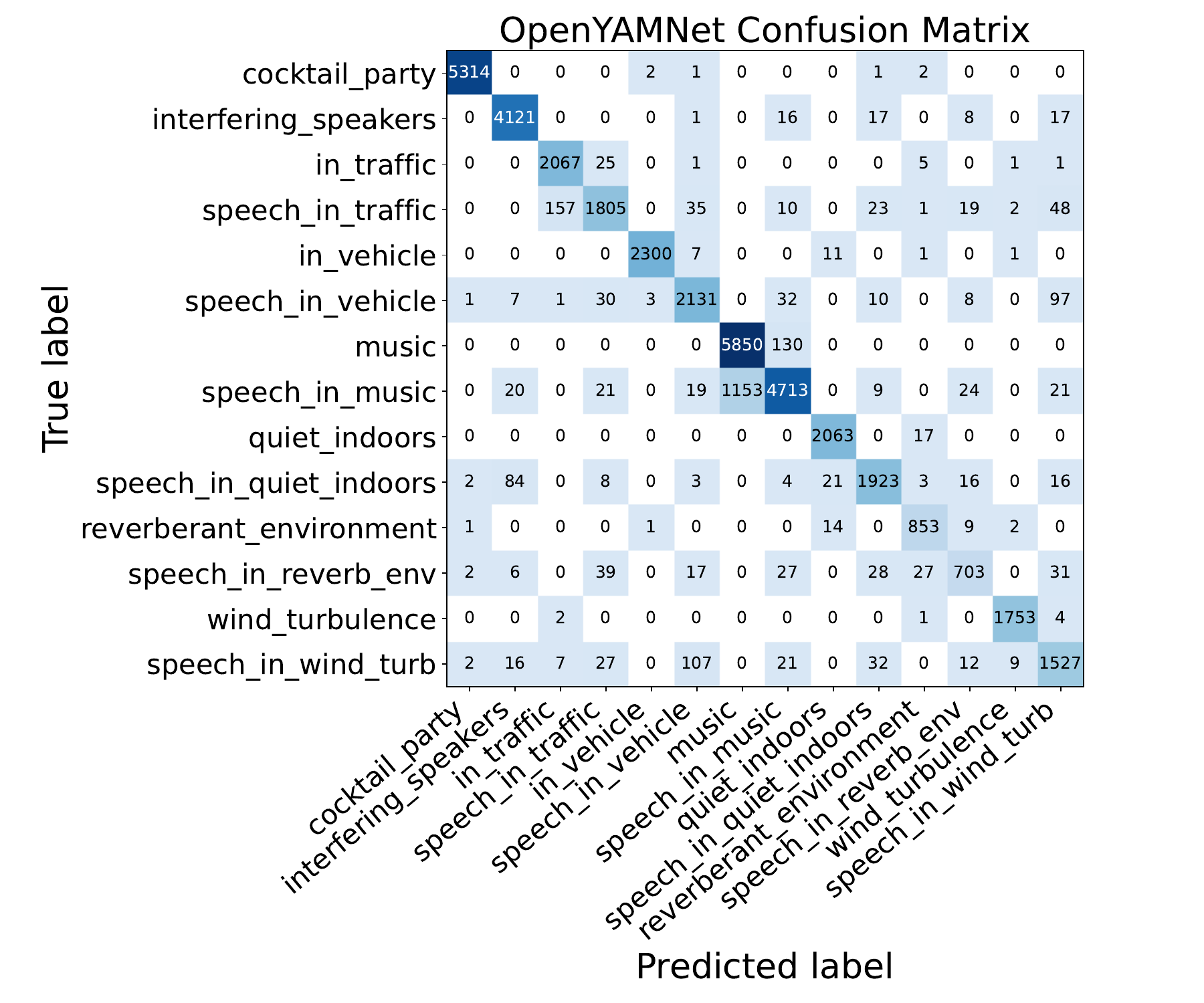}
\caption{OpenYAMNet confusion matrix. An accuracy of 0.93 was achieved on the test set. Each 960ms window contributed a prediction outcome. A higher number along the diagonals indicates better results.}
\label{fig:conf}
\end{figure*}

Precision and recall curves are shown in Figure \ref{fig:precisionrecall}. The confusion matrix is shown in Figure \ref{fig:conf}. The metrics were produced using the test set of AHEAD-DS. Unless otherwise noted, tests were conducted using parameters outlined in the previous sections. Results shown in this section have been rounded to 2 decimal places.

A mAP of 0.86 was achieved. All mAP metrics in this paper were calculated by averaging precision values across thresholds from 0.0 to 1.0, in 0.01 increments and then averaging those over all labels. This gives the true positive, false positive, true negative, and false negative values. A higher mAP indicates better performance.

The confusion matrix indicates an accuracy of 0.93 was achieved. The confusion matrix was produced by computing the confidence of each label. The highest scoring label was compared to the ground truth. A higher number along the diagonals indicates better results.

\subsection{Learning Rate Sensitivity Analysis}
To evaluate learning rate sensitivity, four values were validated: 0.001, 0.0001, 0.00001, and 0.000001. For each learning rate, models were trained three times with different random seeds to obtain the mean and standard deviation (SD) for key metrics: loss, mAP, and accuracy on the validation set. The previously described decay-on-plateau learning rate scheduler was used, and model weights were initialised via transfer learning with weights pre-trained on the AudioSet dataset.

The results showed that learning rates of 0.0001, 0.00001, and 0.000001 all produced very similar performance, with a mean loss of 0.11, mean mAP ranging from 0.85 to 0.88, and a mean accuracy of 0.94. The low SD, which rounded to zero for several metrics, highlighted the consistency of these results. The learning rate of 0.001 performed poorly with a mean validation loss of 2.0, mean validation mAP of 0.5, and Mean Validation Accuracy of 0.14. The geometric mean of the three well performing learning rates, 0.00001 was chosen as the default learning rate.

\begin{table*}
\centering
\captionsetup{justification=centering}
\caption{The mean and standard deviations for loss, mAP, and accuracy over three iterations of each learning rate on the validation set.}
\label{tab:learning}
\begin{tabularx}{0.9\textwidth}{p{0.2\textwidth}p{0.2\textwidth}p{0.2\textwidth}p{0.2\textwidth}}
\hline
\textbf{Learning Rate} & \textbf{Mean Validation Loss $\pm$ SD} & \textbf{Mean Validation mAP $\pm$ SD} & \textbf{Mean Validation Accuracy $\pm$ SD} \\
\hline
0.001 & 2.00 $\pm$ 1.09 & 0.50 $\pm$ 0.27 & 0.14 $\pm$ 0.01 \\
0.0001 & 0.11 $\pm$ 0.00 & 0.85 $\pm$ 0.01 & 0.94 $\pm$ 0.00 \\
0.00001 & 0.11 $\pm$ 0.00 & 0.88 $\pm$ 0.00 & 0.94 $\pm$ 0.01 \\
0.000001 & 0.11 $\pm$ 0.00 & 0.88 $\pm$ 0.00 & 0.94 $\pm$ 0.00 \\
\hline
\end{tabularx}
\end{table*}

\subsection{Ablation Study}
The results of an ablation study validating several configurations are shown in Table \ref{tab:config}. Each metric was calculated by running 10 iterations using bootstrapping, where 200 samples were chosen with replacement from the validation set. The result under the default configuration achieved a mean validation loss, mAP, and accuracy of 0.11, 0.88, and 0.95 respectively. The default configuration had label smoothing enabled, augmentation enabled, all layers were trainable, and weight initialisation used weights pre-trained on AudioSet. The learning rate was set to 0.00001 for all experiments in this section.

The regularising techniques of label smoothing and augmentation were validated. The model produced results which were very close to the models which did use these techniques, approximately within 0.01 mAP. Augmentation by applying a larger gain range of $\pm$24 dB, or use of weighted class training produced a slightly lower mAP, 0.85, but similar accuracy. The lowest validation loss of 0.01 was achieved when label smoothing was disabled. 

When only the last layer was trainable, the ML model produced a mean validation loss, mAP, and accuracy of 0.18, 0.91, and 0.80 respectively. When using random weight initialisation, the ML model produced a validation loss, mAP, and accuracy of 0.53, 0.60, and 0.15 respectively. Using the TAU Urban Audio Visual Scenes 2021 dataset for pre-training produced similar results to random initialisation, a mean validation loss, mAP, and accuracy of 0.53, 0.62, and 0.15 respectively.

Normalising audio to a fixed root mean square (RMS) produced a noticeable drop in mAP and accuracy, 0.75 and 0.84 respectively. Similarly automatic gain control (AGC) produced a noticeable drop in mAP and accuracy, 0.77 and 0.85 respectively.

\begin{table*}
\centering
\captionsetup{justification=centering}
\caption{Ablation study, the mean and standard deviations for loss, mAP, and accuracy for several configurations on the validation set over 10 iterations.}
\begin{tabularx}{0.9\textwidth}{p{0.25\textwidth}p{0.18\textwidth}p{0.18\textwidth}p{0.2\textwidth}}
\hline
\textbf{Configuration} &\textbf{Mean Validation Loss $\pm$ SD} &\textbf{Mean Validation mAP $\pm$ SD} &\textbf{Mean Validation Accuracy $\pm$ SD} \\
\hline
Default & 0.11 $\pm$ 0.00 & 0.88 $\pm$ 0.01 & 0.95 $\pm$ 0.01 \\
No Label Smoothing & 0.01 $\pm$ 0.00 & 0.88 $\pm$ 0.01 & 0.95 $\pm$ 0.01 \\
No Augmentation & 0.11 $\pm$ 0.00 & 0.87 $\pm$ 0.01 & 0.94 $\pm$ 0.01 \\
Unfreeze Only Last Layer & 0.18 $\pm$ 0.01 & 0.91 $\pm$ 0.00 & 0.80 $\pm$ 0.01 \\
Random Weight Initialisation & 0.53 $\pm$ 0.01 & 0.60 $\pm$ 0.00 & 0.15 $\pm$ 0.02 \\
TAU pre-training & 0.53 $\pm$ 0.01 & 0.62 $\pm$ 0.00 & 0.15 $\pm$ 0.02 \\
Augmentation $\pm$24 dB & 0.11 $\pm$ 0.00 & 0.85 $\pm$ 0.01 & 0.94 $\pm$ 0.01 \\
Weighted Classes & 0.11 $\pm$ 0.00 & 0.85 $\pm$ 0.01 & 0.94 $\pm$ 0.01 \\
Normalised RMS=0.2 & 0.17 $\pm$ 0.01 & 0.75 $\pm$ 0.01 & 0.84 $\pm$ 0.01 \\
Block AGC RMS=0.2 & 0.17 $\pm$ 0.01 & 0.77 $\pm$ 0.01 & 0.85 $\pm$ 0.01 \\
\hline
\end{tabularx}
\label{tab:config}
\end{table*}

\subsection{2-Fold Cross Validation}
A two fold cross validation was implemented, the results are shown in Table \ref{tab:kfold}. Each fold was further divided so that ten percent of the samples within the fold were used as a hold-out dataset during training. This was necessary as the training loop expects the presence of a validation set. This created a final 45/5/50 train/validation/test split, where 45/5 belong to one fold for training and validation, and 50 belongs to the other fold for testing.

The results show very similar performance between 2-fold cross validation and the 70/10/20 data split. Both methods achieved a loss of 0.11, mAP of 0.86, and accuracy of 0.93.

\begin{table}
\centering
\captionsetup{justification=centering}
\caption{The loss, mAP, and accuracy for 2-fold cross validation using AHEAD-DS.}
\label{tab:kfold}
\begin{tabularx}{0.5\textwidth}{p{0.2\textwidth}p{0.1\textwidth}p{0.1\textwidth}p{0.1\textwidth}}
\hline
\textbf{Fold} & \textbf{mAP} & \textbf{Accuracy} \\
\hline
Fold 1 train/Fold 2 test & 0.86 & 0.93 \\
Fold 2 train/Fold 1 test & 0.86 & 0.93 \\
Mean & 0.86 & 0.93 \\
\hline
\end{tabularx}
\end{table}

\subsection{Gain Sensitivity Analysis}
\begin{figure*}
\centering
\captionsetup{justification=centering}

\begin{subfigure}[t]{0.45\textwidth}
\includegraphics[width=\textwidth]{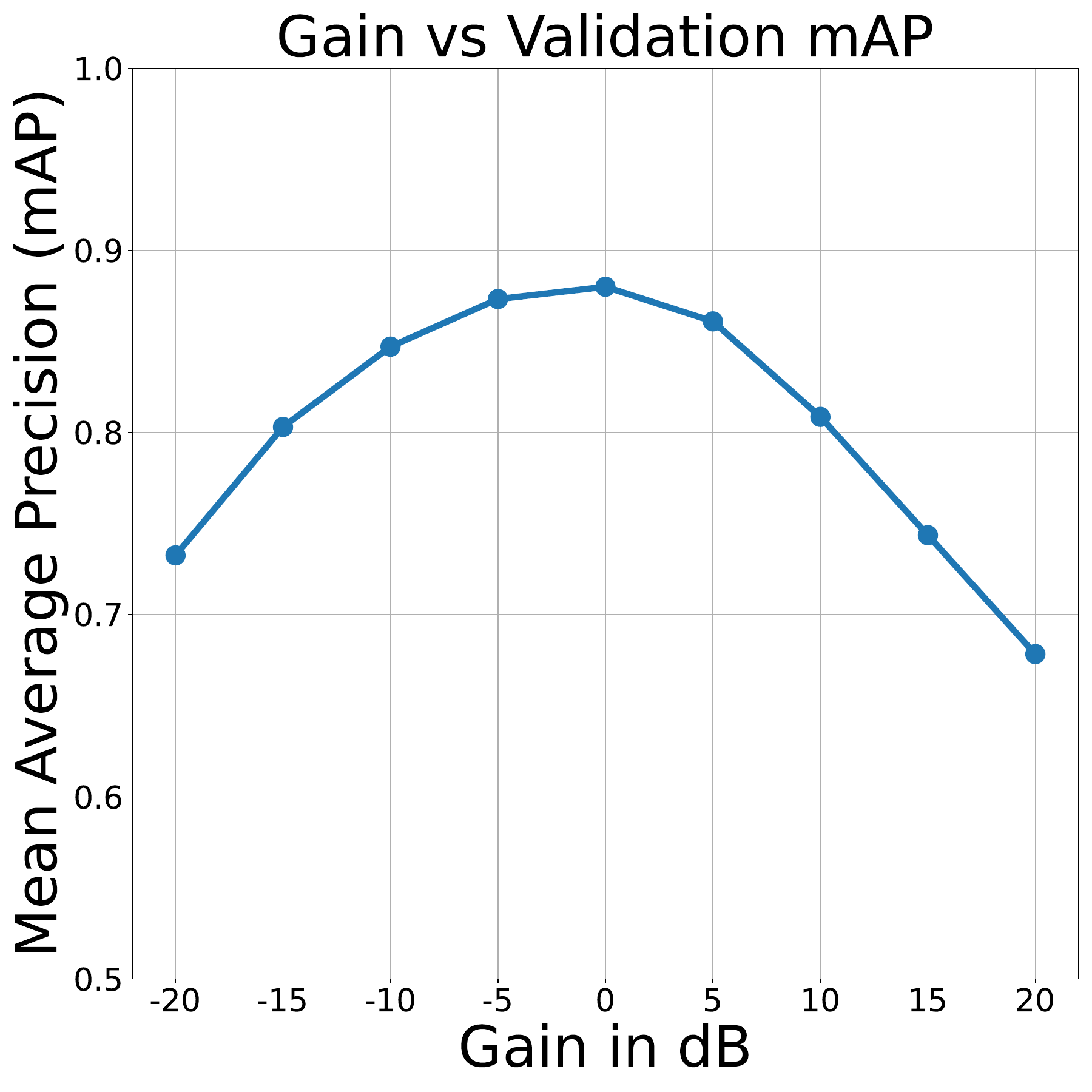}
\caption{Gain vs validation mAP.  Higher values are better.}
\end{subfigure}
\begin{subfigure}[t]{0.45\textwidth}
\includegraphics[width=\textwidth]{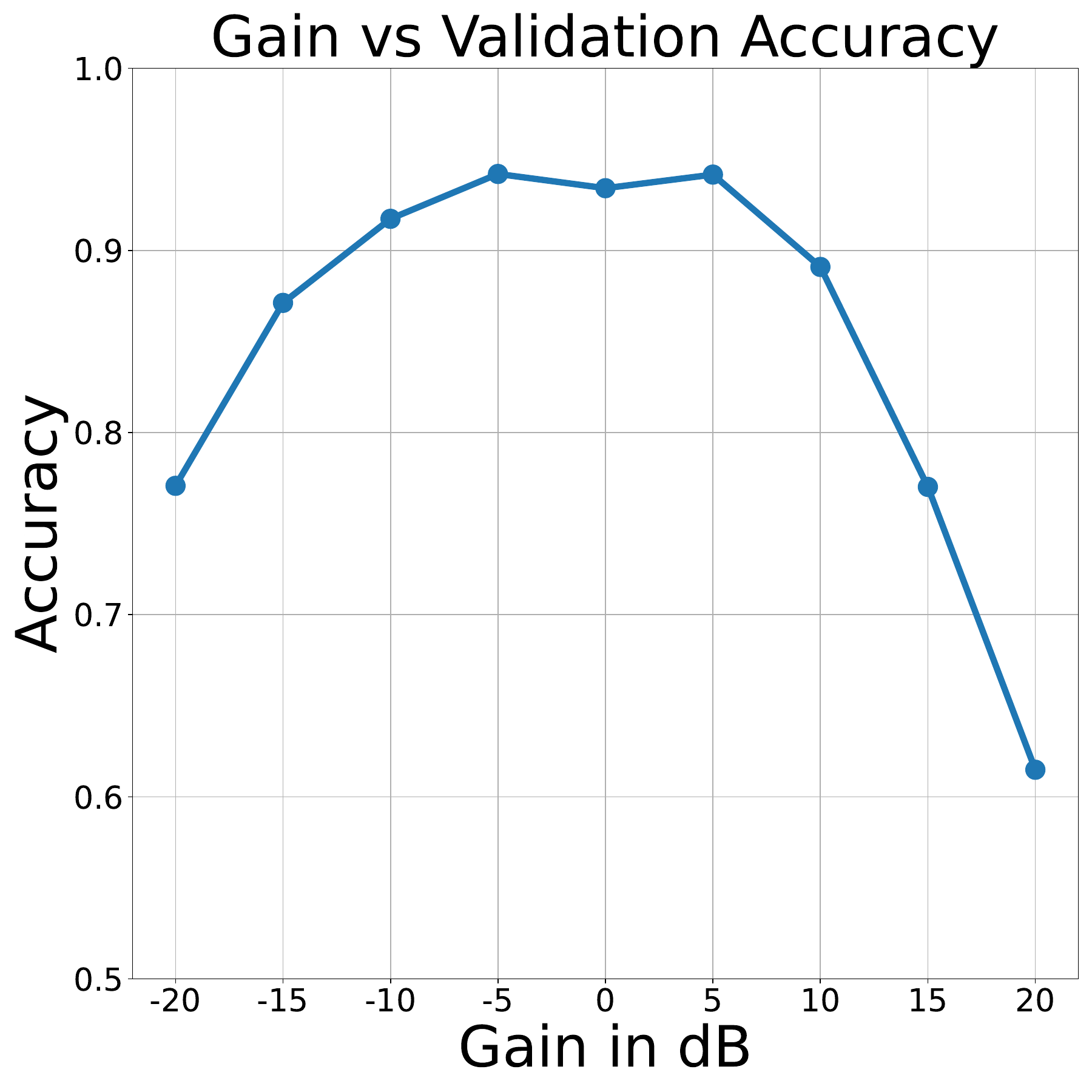}
\caption{Gain vs validation accuracy. Higher values are better.}
\end{subfigure}

\caption{The validation mAP and accuracy at different gain levels.}
\label{fig:gain}
\end{figure*}

Tests were performed at different input gains. The validation mAP and accuracy are shown in Figure \ref{fig:gain}. Validation was performed with a gain of -20 to +20 dB in 5 dB increments. Gain was applied to both speech and background noise, so SNR did not change.Close performance was observed in validation mAP and accuracy when testing with -5, 0, and 5 dB gain. A higher mAP was achieved at 0 dB due to fewer false positives, but the accuracy was lower compared to -5 and 5 dB due to more false negatives.

\subsection{Signal To Noise Ratio Analysis}
\begin{figure*}
\centering
\captionsetup{justification=centering}

\begin{subfigure}[t]{0.47\textwidth}
\includegraphics[width=\textwidth]{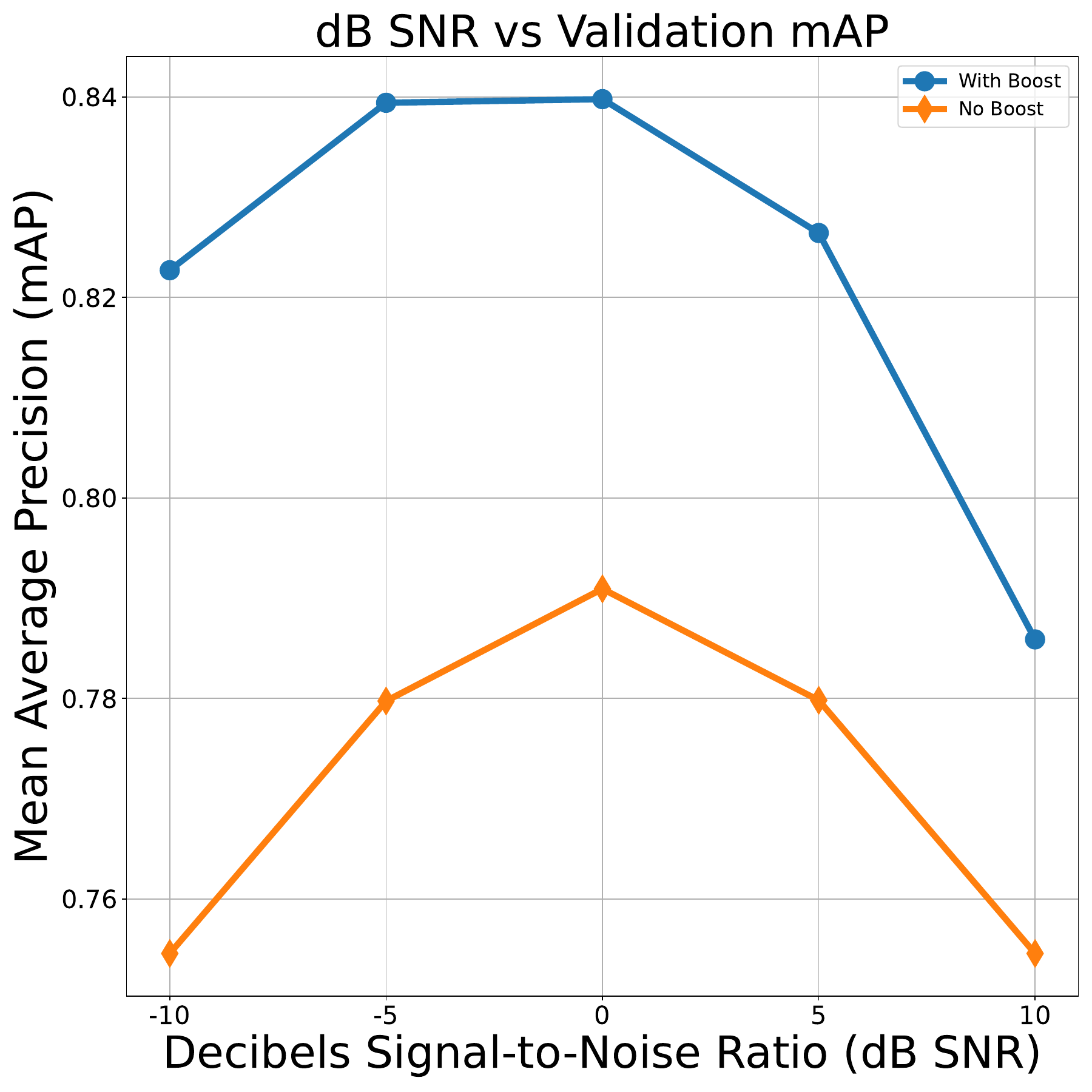}
\caption{SNR vs validation mAP.  Higher values are better.}
\end{subfigure}
\begin{subfigure}[t]{0.47\textwidth}
\includegraphics[width=\textwidth]{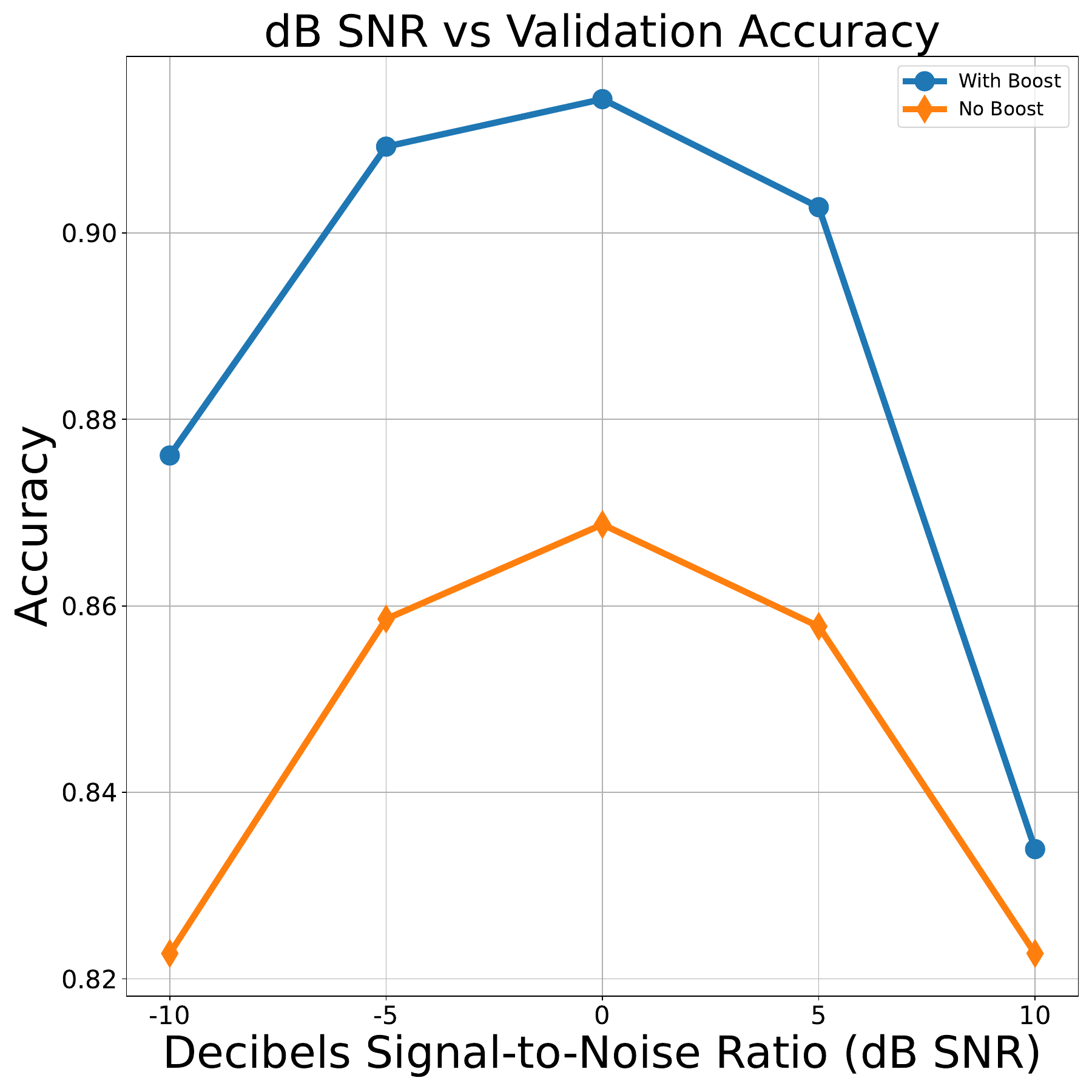}
\caption{SNR vs validation accuracy. Higher values are better.}
\end{subfigure}

\caption{The validation mAP, and accuracy for a series of SNRs.}
\label{fig:snrboost}
\end{figure*}

The goal of these tests were to determine the impact of the mixing and level boosts outlined in Algorithm \ref{alg:mix}. The results are shown in Table \ref{fig:snrboost}.

In the original AHEAD-DS, a total of 2656 of the 6980 training files, 381 of the 1001 validation files, and 756 of the 1987 test files were processed using Algorithm \ref{alg:mix}. The default configuration of hyperparameters were used during training, while validation speech and environment sounds were mixed at a fixed SNRs rather than cycled through the range $-10, -5, 0, 5, 10$ dB, both with and without the level boost.

Best performance was achieved when the validation parameters were closest to those used in training. A $0$ dB SNR with a level boost achieved the best performance with mAP and accuracy at 0.84 and 0.91. Further drops in performance were observed when the SNR was further from $0$ dB. There was a larger drop in performance without the level boost, at $0$ dB SNR the mAP and accuracy were 0.80 and 0.87.

\subsection{Best Per Class Threshold}
The best per class thresholds on the validation set are shown in Table \ref{tab:perclass}. The best threshold was calculated by finding the threshold which produced the largest value when precision is multiplied by recall.

\begin{table}
\centering
\captionsetup{justification=centering}
\caption{Best per class threshold on the validation set.}
\label{tab:perclass}
\begin{tabularx}{0.5\textwidth}{p{0.2\textwidth}p{0.07\textwidth}p{0.07\textwidth}p{0.07\textwidth}}
\hline
\textbf{Class} & \textbf{Threshold} & \textbf{Precision} & \textbf{Recall} \\
\hline
cocktail\_party & 0.41 & 1.00 & 1.00 \\
interfering\_speakers & 0.77 & 0.95 & 0.99 \\
in\_traffic & 0.56 & 0.97 & 0.97 \\
in\_vehicle & 0.58 & 0.98 & 0.99 \\
music & 0.66 & 0.90 & 0.99 \\
quiet\_indoors & 0.35 & 1.00 & 1.00 \\
reverberant\_environment & 0.21 & 0.97 & 0.99 \\
wind\_turbulence & 0.79 & 0.99 & 0.99 \\
speech\_in\_traffic & 0.71 & 0.95 & 0.90 \\
speech\_in\_vehicle & 0.60 & 0.91 & 0.93 \\
speech\_in\_music & 0.93 & 0.98 & 0.85 \\
speech\_in\_quiet\_indoors & 0.59 & 0.95 & 0.91 \\
speech\_in\_reverb\_env & 0.75 & 0.93 & 0.90 \\
speech\_in\_wind\_turb & 0.80 & 0.90 & 0.87 \\
\hline
\end{tabularx}
\end{table}

\subsection{Context Length}
\begin{figure*}
\centering
\captionsetup{justification=centering}

\begin{subfigure}[t]{0.49\textwidth}
\includegraphics[width=\textwidth]{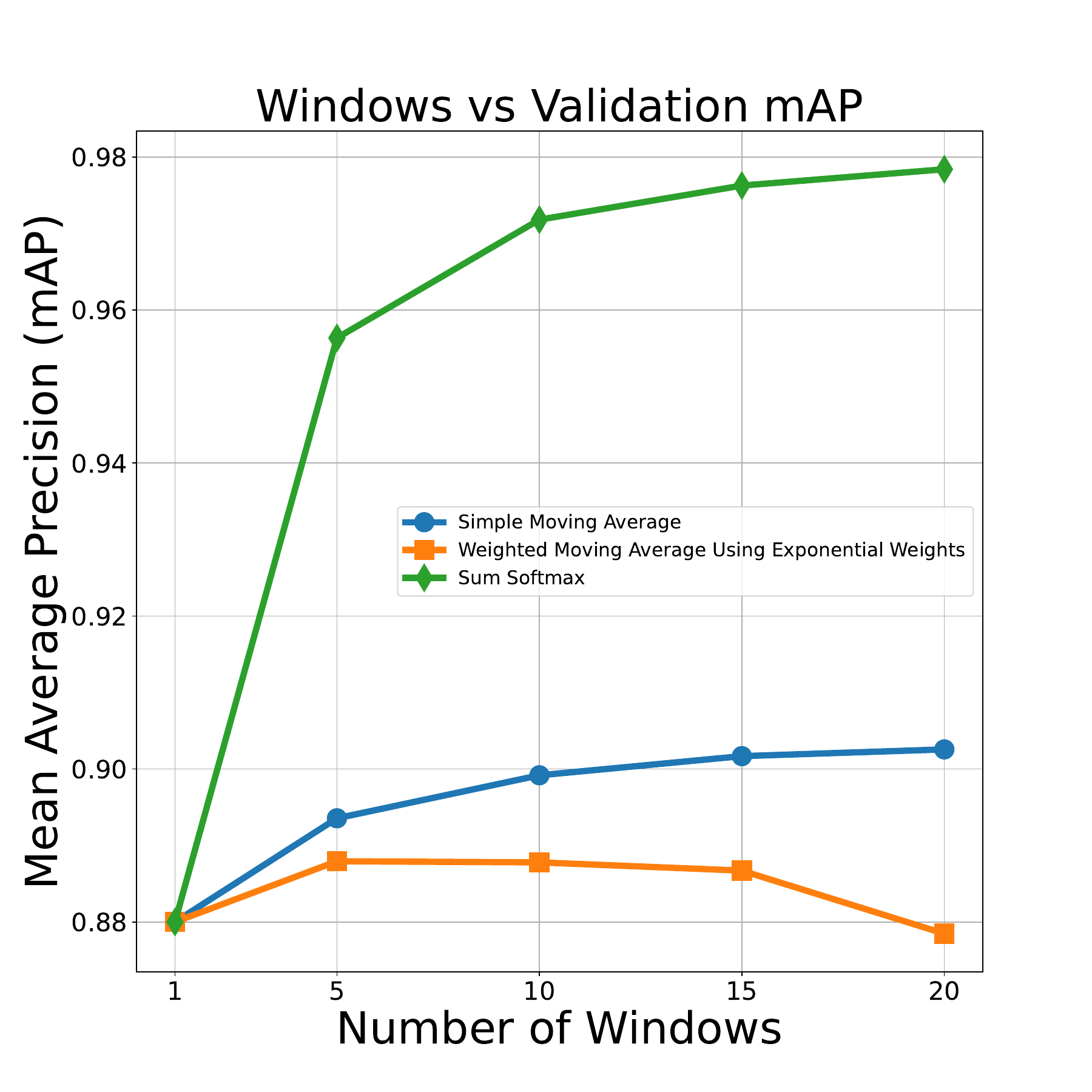}
\caption{Number of windows vs validation mAP.  Higher values are better.}
\end{subfigure}
\begin{subfigure}[t]{0.49\textwidth}
\includegraphics[width=\textwidth]{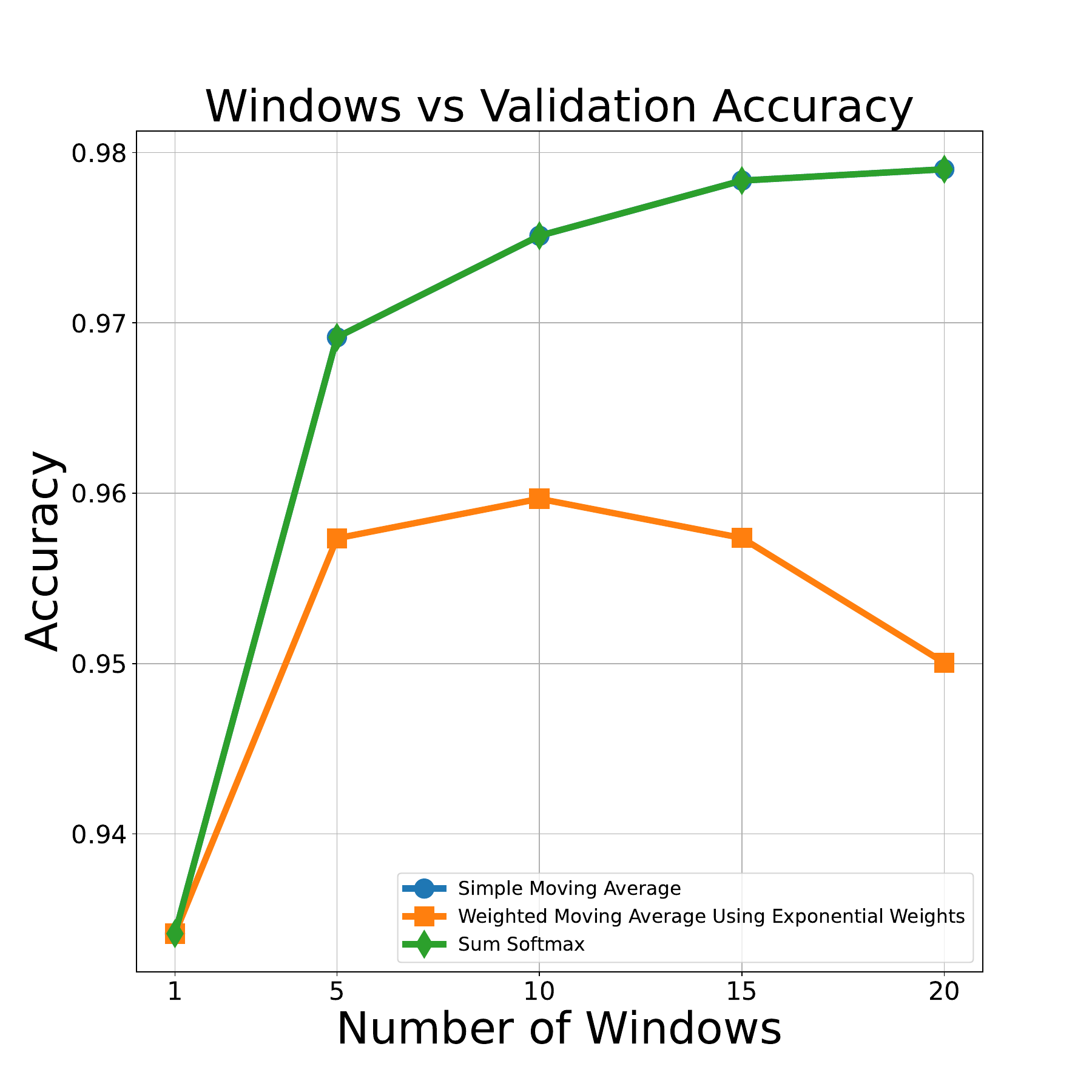}
\caption{Number of windows vs validation accuracy. Higher values are better. Simple moving average and sum and softmax produce the same accuracy.}
\end{subfigure}

\caption{Context length: The validation mAP and accuracy using a different number of windows.}
\label{fig:window}
\end{figure*}

To extend the context beyond a single 960 millisecond window, results from a sequence of consecutive windows were aggregated. The aggregated results, as shown in Table \ref{fig:window}, were evaluated on the validation set. Sequence lengths of 5, 10, 15, and 20 windows were tested. For aggregation, three methods were used: a simple moving average (SMA), a weighted moving average (WMA) with exponential weights, and a summation followed by a softmax function across classes. The exponential weights for the WMA were set as a series of powers of $e$, specifically $e^{0}, e^{1}, … , e^{n}$, where $n$ represents the number of windows in the sequence. Extending the context window beyond 960 millisecond windows improves performance. SMA and sum and softmax achieve the same accuracy as both methods are order-preserving functions, so the highest scoring class remains the same.

\subsection{16-Bit Model Performance and Speed on Edge Hardware}
For edge computing deployment a 16-bit model was used. The 16-bit model achieved a validation mAP of 0.88 and an accuracy of 0.95. The processing times are shown in Table \ref{tab:speed}. The speed tests were performed on a Google Pixel 3, a phone with modest specifications, released in 2018. Segments of 5, 10, 20, and 30s of audio were recorded on the phone, and the time it takes to process the audio by the model was measured. For each length of time, this was repeated ten times. 

Testing was not performed on streamed audio but on recorded audio. The demonstration application works by having a user press a button to start and then another to stop a recording. The application then buffers the last 5, 10, or 30 seconds of audio. To measure processing time, a timestamp was recorded when the audio is buffered into memory and when the results are available. The difference between these two timestamps, in milliseconds, is the total processing time, which includes the time required to compute the log mel spectrograms.

It takes a mean time of 199.4ms to process 5s of sound with a SD of 10.31ms. There was approximately 50ms of latency to load the model, and approximately a linear increase of 30ms per 1 second of audio. A real-time factor (RTF) of between 0.038 and 0.044 was achieved over the ten iterations when processing 5s of sound. The RTF decreases with increasing length of processed audio. The latency of approximately 50ms is a one-off cost, so long as the variable representing the model remains in scope.

The tests were performed by deploying an Android application to the test Android phone. The application was launched and began buffering a maximum of 30s of audio. Audio from the beginning gets discarded as the most recent recording gets buffered. The user was then able to press a button and process a fixed interval of audio buffered in memory \cite{prismpaper}.

\begin{table*}
\centering
\captionsetup{justification=centering}
\caption{The processing times for 5, 10, 20, and 30s of sound over ten iterations on a Google Pixel 3.}
\begin{tabularx}{0.7\textwidth}{p{0.2\textwidth}p{0.3\textwidth}p{0.2\textwidth}}
\hline
\textbf{Audio length (ms)} & \textbf{Mean Processing Time (ms) $\pm$ SD} & \textbf{RTF Range} \\
\hline
5000 & 199.4 $\pm$ 10.31 & 0.037 to 0.044 \\
10000 & 344.5 $\pm$ 12.29 & 0.032 to 0.036 \\
20000 & 671.3 $\pm$ 22.88 & 0.032 to 0.035 \\
30000 & 959.8 $\pm$ 31.12 & 0.030 to 0.033 \\
\hline
\end{tabularx}
\label{tab:speed}
\end{table*}

\subsection{OpenYAMNet vs Perch vs AST}
OpenYAMNet was compared against Perch and AST. The results are shown in Table \ref{tab:compare}. The results were produced using the test set of AHEAD-DS. The two models were trained and validated on the embeddings from Perch and AST using the training and validation sets of AHEAD-DS respectively. Since both Perch and AST require an input length of 160000 (10 seconds), the results from OpenYAMNet were aggregated using the aforementioned sum and softmax procedure, across all 20 windows in each recording. Furthermore, because the models were only trained on the embeddings of Perch and AST, the version of OpenYAMNet used in this test had only the last layer trainable.

\begin{table}
\centering
\captionsetup{justification=centering}
\caption{Comparison of OpenYAMNet against Perch and AST using AHEAD-DS.}
\begin{tabularx}{0.5\textwidth}{p{0.1\textwidth}p{0.1\textwidth}p{0.1\textwidth}p{0.12\textwidth}}
\hline
\textbf{Model} & \textbf{Parameters} & \textbf{Test mAP} & \textbf{Test Accuracy} \\
\hline
OpenYAMNet & 3.7 million & 0.84 & 0.93 \\
Perch & 7.8 million & 0.80 & 0.89 \\
AST & 86.0 million & 0.93 & 0.97 \\
\hline
\end{tabularx}
\label{tab:compare}
\end{table}

\section{Discussion}
This section discusses the results of the tests from the previous section. This includes the implications of the findings, a comparison to existing work, limitations of the current solution, and next steps for the future direction of the research.

\subsection{Summary of Findings}
The experiments demonstrate that OpenYAMNet achieves promising performance, with a test mAP of 0.86 and an accuracy of 0.93. The workflow and toolchain created for training and deploying OpenYAMNet to edge devices is useful for future research in this area.

There were a few classes which were misidentified a notable number of times during testing. The class \emph{speech in music} and \emph{music} were often misidentified during pauses in speech and only music could be heard. The same issue occurs for \emph{speech in traffic} and \emph{in traffic}. The class \emph{speech in vehicle} was sometimes misidentified as \emph{speech in wind turbulence} because the sound of rushing air entering the vehicle can often be heard.

The study highlights two critical findings that are essential for OpenYAMNet to perform well. One is the necessity of transfer learning from the pre-trained YAMNet model on AudioSet to achieve good performance. The other is the necessity of proper adjustment of sound levels to ensure optimal performance.

Two ways of applying transfer learning were tested with OpenYAMNet. One method is to freeze all except the last layer, alternatively parameters in all layers are trainable. The \emph{Ablation Study} showed that training all layers yielded better results than training only the last layer, which significantly impaired model performance. Another finding is that the TAU dataset is not a sufficient pre-training substitute for AudioSet. At approximately twelve thousand audio clips compared to the approximately two million in AudioSet, the TAU dataset is substantially smaller. The tradeoff of using transfer learning is that several hyperparameters become fixed. Changing a hyperparameter impacts all parameters dependent on the said hyperparameter. A new model will need to be initialised and re-trained from scratch using AudioSet, before transfer learning can be applied.

A learning rate of 0.001 was found to perform very poorly. The resulting ML model was likely not practically useful. This was likely because its large value caused the model parameters to overshoot optimal values.

The \emph{Ablation Study} section tested several hyperparameter configurations. The two regularising techniques of label smoothing and augmentation did not have a major impact on the validation mAP and accuracy. The two techniques were kept enabled as the performance drop was minimal. The one noteworthy result was the validation loss of 0.01 when label smoothing was disabled. The result is expected as smoothing lowers the value of the true class while increasing the values of the negative classes during training. When a more extreme augmentation range using $\pm$24 dB was validated, there was a slight dip in performance. Since the gain is uniformly distributed between $\pm$24 dB, the distribution introduces occasional outliers at or near $\pm$24 dB leading to a performance drop. The use of weighted classes also caused a slight drop in performance. Focal loss already compensates for class imbalance. Therefore, the use of weighted classes does not seem necessary. AGC and normalisation caused a drop in performance. Both techniques discard information on the relative differences in level between environments.

A 2-fold cross validation test was conducted to determine if the previous experiment results were an artifact of the 70/10/20 training/validation/testing split. The 2-fold cross validation achieved equivalent mAP and accuracy to the test set. This is a strong indicator that the results produced using the 70/10/20 split are not an artifact of the split. The 70/10/20 split was based on a rule of thumb.

In gain sensitivity tests, deviations in sound levels outside the -5, 0, and 5 dB range led to diminished performance. This indicates adjustment of the levels is necessary to achieve optimal performance when applying the model to new datasets.

In SNR analysis there was a drop in performance without the level boost from Algorithm \ref{alg:mix}. This was due to a large drop in metrics for \emph{speech in quiet indoors}. Without the level boost, the addition of speech did not significantly change the signal after quantisation. The reason to detect speech at -10 dB SNR was to test the possibility of detecting speech, though not intelligibility, in edge cases that would challenge a human with good hearing.

The validation results indicate it is possible to increase precision and accuracy by setting per class thresholds.

Extending beyond a single window smooths out the results, reducing the effects of occasional errors leading to overall higher mAP and accuracy. When using WMA with exponential weights and a context length of 20 windows, the mAP is lower than a single window and the accuracy also drops compared to 15 windows. Each 10 second audio sample is divided into 20 overlapping windows, the weight of the $20th$ window is $e^{20}$, which dominates the results of the previous 19 windows. If the $20th$ windows is misclassified, the entire audio sample will likely be misclassified, this is detrimental to performance. Sum and softmax show the largest increase to mAP. This is because softmax further boosts the score of the class with highest confidence relative to the scores of other classes.

The real-time sound-based scene recognition capabilities of OpenYAMNet on edge devices were successfully demonstrated on an edge device. With a RTF well below 1.0, this indicates the feasibility of deploying such models for real-time applications on resource constrained devices. There are more performant edge devices compared to our test system of a Google Pixel 3, even lower latency and processing time can be achieved with newer devices. The 16-Bit model deployed to the edge system achieved a 0.88 mAP, and 0.93 accuracy. The results indicate the 16-Bit model is practically indistinguishable from the 32-bit model.

\subsection{Comparison with Previous Work}
AHEAD-DS is based on HEAR-DS, which provides a solid foundation as the recordings of environment sounds have been open sourced with a permissive licence. One issue with HEAR-DS is that the dataset is not in a fixed, ready-to-use state. This is because some of the published data is unmixed, and some pseudo-random procedures were used during the production of the HEAR-DS dataset. Since the code has not been published, these procedures cannot be fully reproduced. The advantage of AHEAD-DS is that the dataset is in a ready-to-use state. This is important because it removes a variable when comparing ML models. The speech and environment sounds have been released in both mixed and unmixed forms, eliminating ambiguity.

OpenYAMNet was compared against Perch and AST. It outperforms Perch in both performance and efficiency by achieving a higher mAP, accuracy, while using fewer parameters, 0.84, 0.93, and 3.7 million respectively compared to Perch's 0.80, 0.89, and 7.8 million. Although AST surpasses OpenYAMNet with the highest mAP and accuracy at 0.93 and 0.97, it is less efficient as it uses 86.0 million parameters.

Compared to the original YAMNet public release, OpenYAMNet is a fully open source re-implementation of YAMNet with a complete workflow, so it can be tuned with transfer learning. YAMNet was published without training code and can only be used as a black box.

\subsection{Limitations}
According to the authors of HEAR-DS, the 14 classes have been chosen because these classes are relevant to auditory scene recognition for hearing devices. These classes were chosen by audiologists and Audifon, a manufacturer of hearing aids. No further rationale was provided by the authors of HEAR-DS for these classes. Since AHEAD-DS adopts the same classes, this is a limitation inherited by AHEAD-DS.

Compared to datasets like AudioSet, AHEAD-DS is still relatively small, which is a disadvantage. As shown in the \emph{Ablation Study}, training from scratch using AHEAD-DS leads to suboptimal results, scoring a validation loss of 0.53. Transfer learning is essential, as the model performs much better with a mAP of 0.86, when fine tuning weights pre-trained on AudioSet.

The use of transfer learning comes with its own tradeoffs. Hyperparameters are locked in due to transfer learning. This makes it resource intensive to change hyperparameters such as the window length. Extending it beyond 960ms will require the model be trained from scratch using AudioSet. It is possible to work around the fixed length by aggregating the results from multiple windows and applying a softmax operation. 

The tests of gain sensitivity and SNR analysis indicate adjustment of the levels is necessary to achieve optimal performance. Recordings using new hardware configurations must be standardised, this only needs to be done once per device, to ensure recordings of conversations using the new device have similar levels to CHiME 6 Dev.

A notable limitation is the dependency on TensorFlow. OpenYAMNet is built using the TensorFlow ML library, as it is necessary to maintain compatibility with YAMNet weights, required for transfer learning. TensorFlow has GPU compatibility problems, disabling the use of a GPU for quick model training and complicating the codebase.

\subsection{Next Steps}
Future work can target three areas: the dataset, the model, and implementation details. The dataset is still relatively small compared to the likes of AudioSet. Therefore, the next step is to collect more data for a new version of AHEAD-DS. More sound clips, more varied sound environments, and more varied recording equipment configurations ensures models trained using a future version of AHEAD-DS are more generalisable.

The dataset can be improved by choosing scene labels in a systematic way. In this paper class labels are adopted labels from an existing study. There is a lack of rigorous studies to explain why these labels are good choices. Examples such as \emph{speech in noise} seem very generic and \emph{one talker on subway platform} seem oddly specific. The balance between labels being too specific or too generic are yet to be determined. This is important as good scene labels are crucial in hearing research, especially when conducting Ecological Momentary Assessment (EMA) \cite{holube2020ecological}. EMAs are common in hearing research and involve surveying participants in real-time, and scene recognition models can automate this process, saving time and reducing the participants' cognitive load. However, if these labels are ambiguous, inconsistent, or too broad/narrow, model accuracy will suffer, negating the benefits of automation and potentially increasing the burden on participants.

It may be possible to improve the results by making better use of the limited data in AHEAD-DS. Speech and environment sounds can be left unmixed but standardised in the training set. More varied training data can be generated if mixed on-the-fly. So long as the test data is not mixed on-the-fly, results still remain comparable between test runs.

It may be possible to achieve better results by using strong annotations with AHEAD-DS. Unlike weak annotations, strong annotations precisely mark the start and end times of each sound.

Things that would improve the model implementation include rebuilding the model by first cleanly separating feature computation and model code. This will enable further experimentation with other features such as time domain based features. This may enable more effective detection of acoustic properties such as reverberations, which are more salient in the time domain. Rebuilding the model enables the ability to test different architectures such as EfficientNet or ResNet. Newer architectures such as EfficientNet might boost precision and recall.

The model implementation can be improved by using a modern ML library such as JAX or PyTorch. This allows us to avoid the complex legacy code and backward compatibility problems of TensorFlow. It would also significantly speed up training as more modern ML libraries have better compatibility with GPU acceleration.

A battery and thermal impact test of the ML model on various types of edge hardware would be a valuable contribution. This will help determine the feasibility of using the model on more types of highly resource constrained devices such as hearing devices.

\section{Conclusion}
This study addresses key challenges in scene recognition for hearing devices by introducing a refined dataset and machine learning model. AHEAD-DS is a publicly available dataset repacked and refined from several other open source datasets. It is designed for auditory scene recognition for hearing devices and offers a standardised dataset for comparing models with consistent labels and in a ready-to-use state. Complementing this, OpenYAMNet is presented as a robust sound recognition model, it only requires 3.7 million parameters, and optimised for deployment on resource-constrained edge devices like smartphones. It serves as an open-source baseline, including model definitions, training, and testing code. Benchmarking OpenYAMNet on AHEAD-DS yielded a mAP of 0.86 and an accuracy of 0.93. The model's efficiency was demonstrated on a Google Pixel 3 smartphone, processing each second of audio in approximately 30ms with 50ms latency. The established workflow and toolchain for training and deploying such models on edge devices offer a valuable resource for future advancements in this field.

\section*{Acknowledgments}
This work was supported by the Google DFI Catalyst fund and Macquarie University.

\bibliographystyle{IEEEtran}
\bibliography{scene_recognition_paper}

\end{document}